# Additive Manufacturing of Lunar Regolith for Reconfigurable Building Blocks toward Lunar Habitation

Cole McCallum[1], Youwen Liang[1], Nahid Tushar[1], Ben Xu[2], Bo Zhao[2], Hao Zeng[3], Wan Shou[1,*]

[1]Department of Mechanical Engineering, University of Arkansas, Fayetteville, AR 72701, USA
[2]Department of Mechanical and Aerospace Engineering, University of Houston, Houston, TX 77204, USA
[3]Faculty of Engineering and Natural Sciences, Tampere University, P.O. Box 541, FI-33101 Tampere, Finland

*Corresponding Author: wshou@uark.edu

**Abstract**

Utilizing locally available materials is a crucial step towards sustainable planetary habitation. Lunar regolith has gained tremendous interest in additive manufacturing in the past decades. However, due to the constrained manufacturing facilities and materials on the moon, many existing additive manufacturing methods are not suitable for practical on-site manufacturing. Here, we envision that light-based direct sintering of lunar regolith can be a feasible approach. Instead of directly manufacturing large structures, we hypothesize that small-scale, reconfigurable building blocks can be an alternative to form large and complex structures. To verify the feasibility, we conducted laser sintering of lunar regolith simulants as a proof of concept, following a simple theoretical calculation for direct sintering using the light available in space. Different laser processing parameters are investigated to obtain controllable lunar regolith sintering. We further designed Lego-like interlocking bricks that are reconfigurable for different structure assemblies without additional material. Mechanical performance (compressive strength) of sintered cubic blocks is evaluated, showing a peak stress of ~1.5 MPa. We hope this work will inspire other in-space manufacturing techniques and enable low-cost space habitation.

**Keywords:** In-Situ Resource Utilization, Assembly, Selective Laser Sintering

**1. Introduction**

Since the first Apollo missions in the late 1960s, establishing a semi-permanent or permanent manned colony on the lunar surface has been subject to rigorous study and exploration. One of the most severe limitations of lunar colonization is the high cost of transporting structures and materials from Earth, highlighting a critical need for onsite manufacturing in space, exploiting local resources [1,2]. Recently, additive manufacturing (AM, a.k.a. 3D Printing) technologies have been considered promising to realize complex solid structure construction with minimal material waste and post-processing work [3, 4]. Thus, various AM techniques are also actively studied for potential on-site construction on the Moon [5-7], using local resources such as lunar dust and regolith. Nieke et al. demonstrated aerosol deposition of lunar regolith simulant (EAC-1) into thick films [8]. Fateri and Gebhardt reported selective laser melting (SLM) of lunar regolith [9], and decent 3D structures and parts were demonstrated. Thermoplastic powders were also added and sintered together with simulants as binders to enhance the strength [10, 11]. Extrusion-based 3D printing methods are also reported to manufacture structures using lunar regolith [12-14], which may be followed by sintering; however, additional polymer binder and solvents are used to prepare the ink, which might be challenging in space. Vat photopolymerization approach was also reported using lunar regolith simulant paste for AM [15,16]; however, the addition of polymers and preparation of paste make the adoption for in-space manufacturing unrealistic. Among various AM approaches, it is believed that light-based sintering is most promising as solar energy is consistently present in great abundance, unlike other power sources commonly utilized for this purpose on Earth. Based on this assumption, researchers have also developed solar 3D printing, where (simulated) sunlight is directly utilized and concentrated with optics to sinter lunar regolith

[17, 18]. It was demonstrated that the manufacture of glass and mirrors from lunar regolith simulant is feasible [19], which means that critical components can be manufactured on site. Therefore, it is highly possible to use minimal components to build the simplest SLM 3D printer on the Moon.

While industrial laser-based additive manufacturing systems typically deliver energy densities in the range of $10^9$–$10^{11}$ W/m² at the melt pool, the solar flux available on the Moon is approximately 1360 W/m² due to the absence of an atmosphere [20]. By employing optical concentrators such as Fresnel lenses or parabolic mirrors, this flux can be intensified by factors of up to 10,000×, yielding focused power densities of 1–10 MW/m². For example, solar furnaces on Earth, such as NREL's High-Flux Solar Furnace, have achieved focal intensities exceeding 2.5 MW/m² and temperatures above 1800 °C [21]. These levels are sufficient to sinter and even partially melt lunar regolith simulants, as demonstrated in studies where sintering occurred at fluxes of 100–200 kW/m² and temperatures ranging from 900–1200 °C [22, 23]. Although this is still orders of magnitude lower than the intensity achieved by high-power lasers, the energy delivered by concentrated solar flux in space is technically adequate for regolith sintering, provided that longer exposure times or slower scan speeds are acceptable.

Although light-based sintering approaches have been widely studied to manufacture lunar regolith parts, directly manufacturing large structural parts with consistent properties remains challenging, due to the complex light-matter interactions and the variability in the composition and shape of regolith [24]. Thus, we hypothesize that developing interconnectable modular structures at a small scale can be an alternative to assembling large parts and structures. Although previous AM work has demonstrated simple brick designs using lunar regolith [17, 25, 26], a reconfigurable brick structure has not been considered. Buildings on Earth typically use cement to bond bricks;

however, simplifying this bonding mechanism is beneficial for the construction of lunar habitation. Therefore, an approach employing a small number of interchangeable, standardized components is desired. Inspired by traditional mortise and tenon joints and Lego-like structures for construction [27, 28], we explore the feasibility of direct laser sintering of lunar regolith into reconfigurable modular bricks for future construction.

## 2. Materials and Methods

### 2.1. Materials

The exact composition of lunar regolith varies slightly depending on where it's collected, however, samples taken from the Apollo missions across multiple sites always contain the compounds $SiO_2$, $Al_2O_3$, CaO, MgO, FeO, $TiO_2$, and MnO in significant quantities [29, 30]. Among these compounds, $SiO_2$ is the most abundant, although the other oxides present in regolith contribute about 50~60% to the overall mass.

In this paper, the terms "lunar regolith" and "lunar regolith simulant" are used interchangeably. It should be noted that all experiments are performed using LHS-1 Lunar Highlands Simulant (provided by Space Resource Technologies), with a specific composition listed in Table 1. For lunar regolith and any of its simulants, it can not be assumed that values such as the density and heat transfer coefficient remain constant, especially at the high temperatures needed for sintering to occur [31], therefore it will be necessary to experiment with different values to find the conditions under which sintering occurs, but not complete melting of regolith.

**Table 1**. Chemical composition of lunar regolith simulant (provided by the vendor).

| Oxide | $SiO_2$ | $TiO_2$ | $Al_2O_3$ | FeO | MnO | MgO | CaO | $Na_2O$ | $K_2O$ | $P_2O_5$ | LOI |
|---|---|---|---|---|---|---|---|---|---|---|---|
| wt.% | 49.12 | 0.63 | 26.29 | 3.20 | 0.06 | 2.86 | 13.52 | 2.55 | 0.34 | 0.17 | 0.41 |

## 2.2 Process and Design

A small-scale powder bed is constructed for brick fabrication (as shown in Figure 1). A motorized Z stage is placed under an 80 W fiber laser (wavelength of 1064 nm, OMTech). In-plane motion (x and y) is realized by the galvanometer scanner. A metal platform (80 cm by 80 cm) was built and attached to the adjustable z-axis stage. The powder bed is designed to hold a powder thickness of up to 80 mm. By moving down the Z stage (or printing platform), and adding powder with controlled thickness, we can realize layer-by-layer sintering for additive manufacturing of regolith. During our printing, the relative distance between the laser head and the top layer of regolith is fixed. Here, our powder deposition is manually performed through powder pouring, followed by brushing to make the surface even with the surface of the powder bed wall. This ensures that the laser is always at the correct focal length when sintering a new layer of regolith simulant. A frequency of 4,000 kHz was selected and fixed for the experiment. The printing pattern is designed using LightBurn software, which is essentially a 2D pattern. By moving the Z stage, a 3D structure can be programmed and sintered.

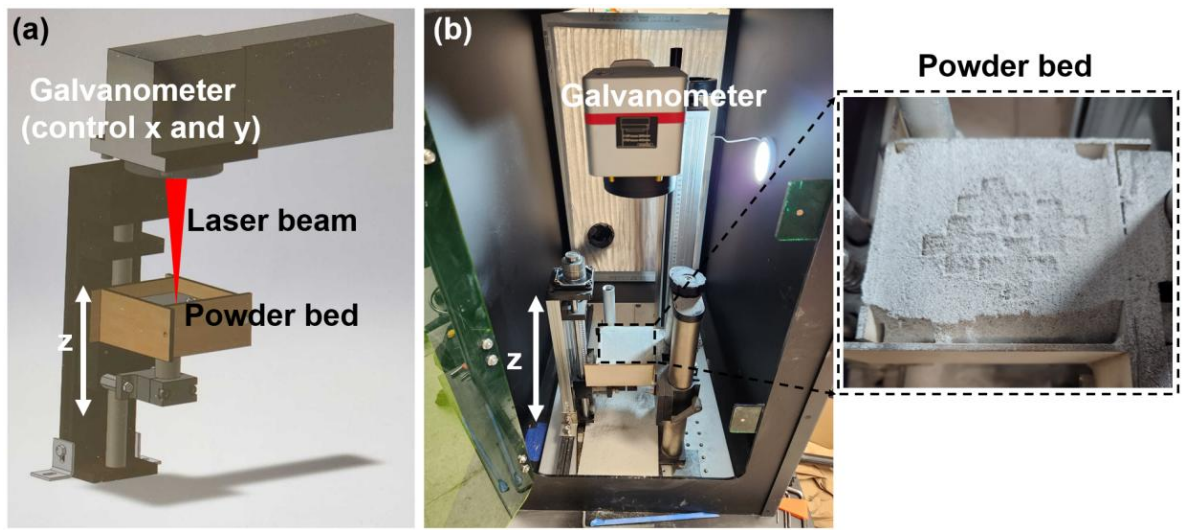

**Figure 1.** The manufacturing system setup. (a) Schematic drawing and (b) photograph.

To test the processing window for sintering (with partial melting), a 0.1 cm by 0.25 cm line is designed with different laser processing parameters (including scanning speed and corresponding energy density). While different hatch spaces (ranging from 0.1 mm to 0.5 mm) were studied, only 0.1 mm gave reasonable sintering/melting with various power and scanning speed combinations. Thus, 0.1 mm was used as a default hatch space. To determine which samples were the most promising for binderless printing, a tweezer was used to delicately remove and store samples. The actual width of each layer was measured by taking the mean width of the sample's impression in the powder bed, whereas the depth was measured directly from the samples.

To realize additive manufacturing for thick and 3D parts, multi-layer sintering (or printing) was further studied. The unit layers exhibiting the most desirable sintering characteristics were recorded, and their sintering parameters were used for further study. To determine the limits of multi-layer sintering, circular layers were sintered from regolith (with different numbers of layers from 1 to 9). The cylinders were then measured to find the relationship between thickness and the number of layers.

After verifying multilayer binderless sintering, the laser parameters that produced the thickest cylinders were recorded and used to build more complex structures. Lego-like structures were fabricated as reconfigurable bricks to test if they can be connected to others to build larger and complex structures. A simple two-part building unit is designed to work as male and female connectors for 2D or 3D structures. This design carried the advantage that such structures could be built without varying the shape of individual layers, making the sintering process comparatively simpler.

## 2.3 Characterizations

Optical images were obtained using a Zeiss microscope with side light illumination. The compressive mechanical test was performed using a Mark 10 tester and M5-300 force gauge. The loading speed of the tester is 1 mm/min, and the force gauge obtained 10 readings per second.

## 3. Results and Discussion

### 3.1 Morphology evolution with different laser processing parameters

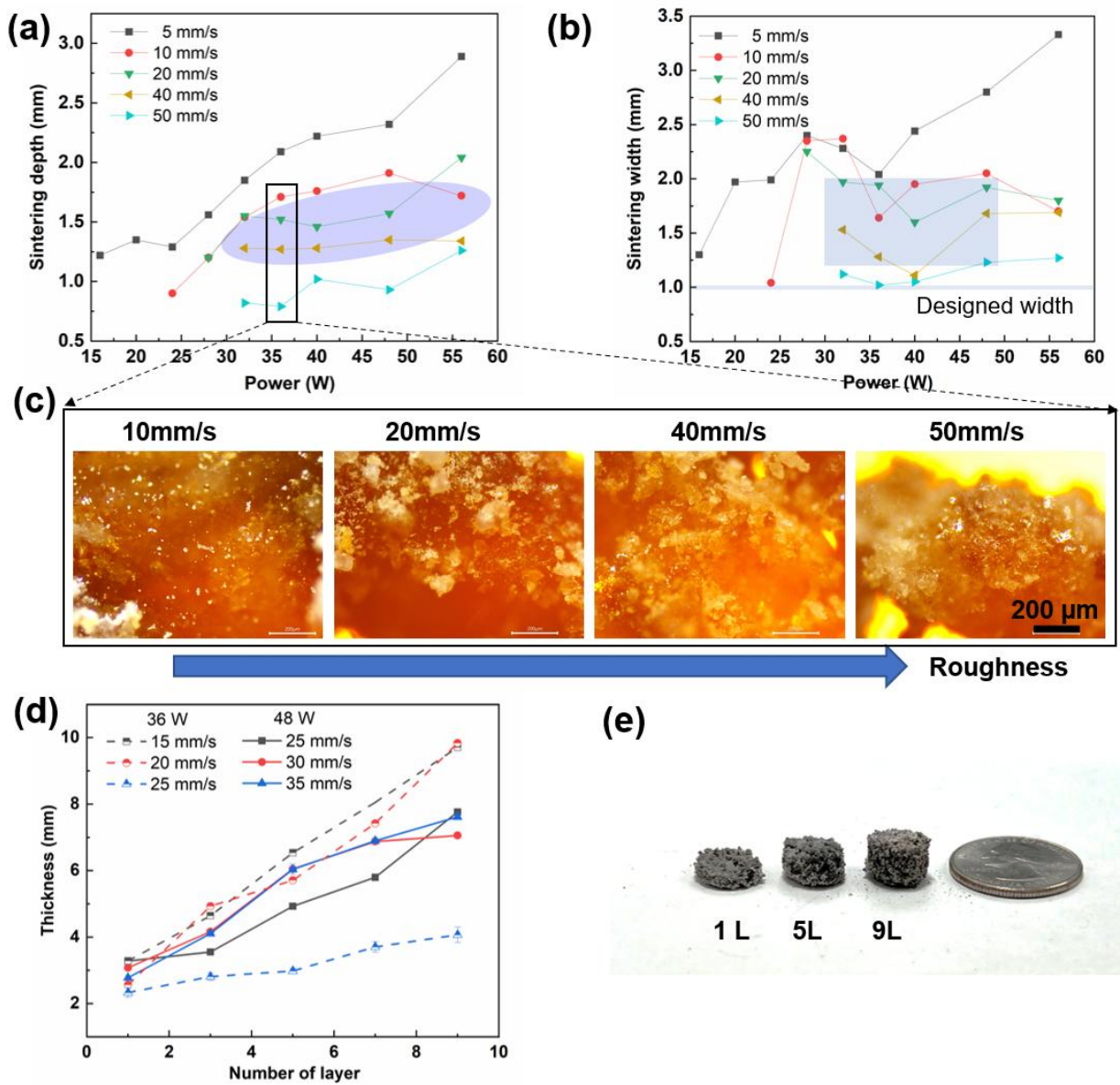

**Figure 2**. Laser processing for structure sintering. (a) Power influence with different scanning speeds on sintering depth. The shadowed area indicates a well-defined sintering window. (b) Power and scanning speed influence on sintering width. (c) Morphology evolution of lunar regolith with different scanning speeds. (d) Laser processing parameters influence on thickness of circular shapes. (e) Photographs of representative cylinder samples with different layers.

To understand the influence of laser processing parameters on the raw lunar regolith, a systematic study was conducted. Two critical measurements, sintering depth and width, were collected using our defined line design (0.1 cm by 0.25 cm). While a wide range of parameter combinations were tested, only the combinations that gave a well-defined line were used for measurement. As shown in Figure 2(a), generally, the sintering depth increases with the increase of laser power. Meanwhile, with the decrease of scanning speed, more heat will be accumulated, and the heat-affected zone will also be deeper. While our designed linewidth is 0.1 cm, the resulting linewidths are always bigger, as shown in Figure 2(b). A similar trend of sintering width is observed, where more thermal effect leads to wider linewidth. However, due to the irregular shape and size distribution of lunar regolith [18], the influence of laser processing parameters on line width is not well-controlled. It should also be noted that melting can also contribute to structure variation. The unit layers were evaluated and sorted into three categories: well-defined, poorly defined, and incomplete sintering based on evaluation under a microscope and by eye inspection. The well-defined processing window is shadowed in Figure 2 (a) and (b), which gives us free-standing samples for consistent measurement. A set of microscope images is shown in Figure 2(c) to show the evolution of the sintered structure with a fixed power of 36 W and different scanning speeds. It can be observed that the surface of the sintered line becomes rough with the increase of the speed. This is closely related to the melting or sintering condition, with a higher energy density ($\sim 4P/(v*\pi*d^2)$) (here P indicates the power, v represents scanning speed, and d represents the beam diameter), the regolith tends to melt and form a smoother surface; while insufficient energy, results in partial melting and rough surface [24, 32].

Based on the results from single-track line testing, multi-layered cylinders were manufactured. A decent linear relationship is observed (Figure 2d), namely, the sample thickness increases with the number of layers. However, it is also noted that there are some variations from layer to layer, which can be caused by the nonuniform distribution of regolith and different thermal conduction conditions due to the change of thickness. In addition, delamination becomes more serious as additional layers are added. Overall, it was found that a lower power of 36 W and speeds of 15-20 mm/s produced cylinders with the greatest depth per layer. Representative cylinders are displayed in Figure 2e, showing a clear thickness increase from 1 layer to 9 layers.

### 3.2 Mechanical Property Evaluation

To determine the use case for binderless sintering, compressive mechanical testing was performed on 1×1×1 cm cubes (Figure 4a) manufactured using the optimized parameters. The corresponding stress-strain curves are shown in Figure 4(b). Here, the samples were tested in different directions: the thickness direction (S1 and S2) and the laser scan direction (S3 and S4). The results show that the sample's modulus and ultimate compressive strength are significantly higher when subjected to loads in line with each layer. However, the maximum strength is about 1.5 MPa, which, while comparable to some previous reports [10, 26, 33], is still not yet suitable for practical application. The measured strength in the laser scan direction is over 2 times higher than in the thickness direction, indicating a high anisotropicity of the mechanical property. The nonuniform distribution and random shape of the granular particles may contribute to this. Also, it should be noted that the granular particles of the simulants can not be spread across the powder bed as smoothly as spherical powder. Therefore, the sintered structures have a rough surface and heterogeneous porosity, which leads to relatively low mechanical properties. Although the current mechanical strength is not sufficient for practical applications, further processing optimization and post-processing can significantly enhance the properties [32].

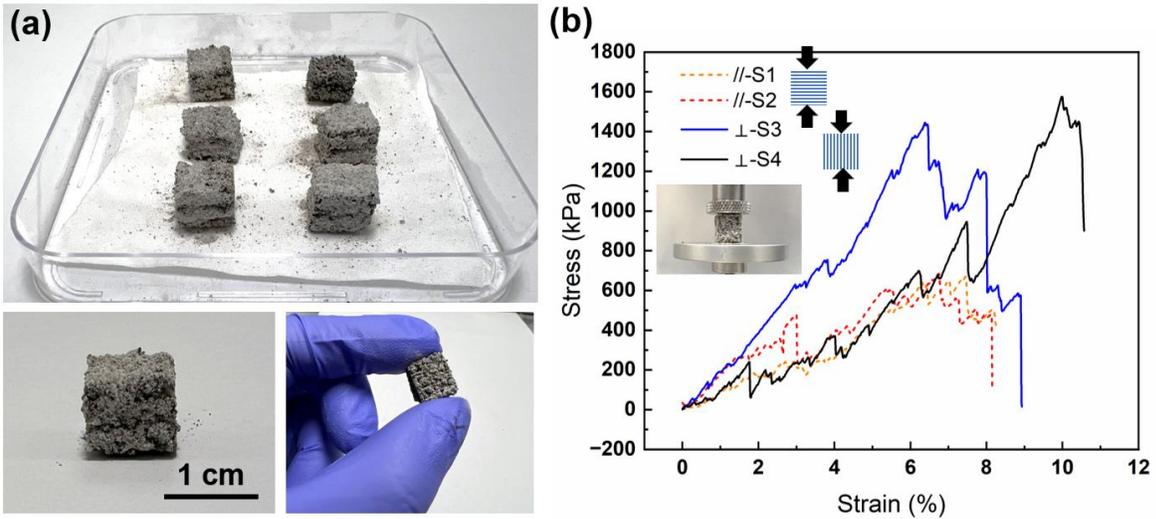

**Figure 3.** Compression test of cubic lunar regolith samples. (a) Photographs of the manufactured test samples. (b) Stress-strain curves of four representative samples measured from different directions. (Samples S1 and S2 were compressed from the printing thickness direction; samples S3 and S4 were compressed from the laser scan direction.)

### 3.3 Additive manufacturing of reconfigurable brick block

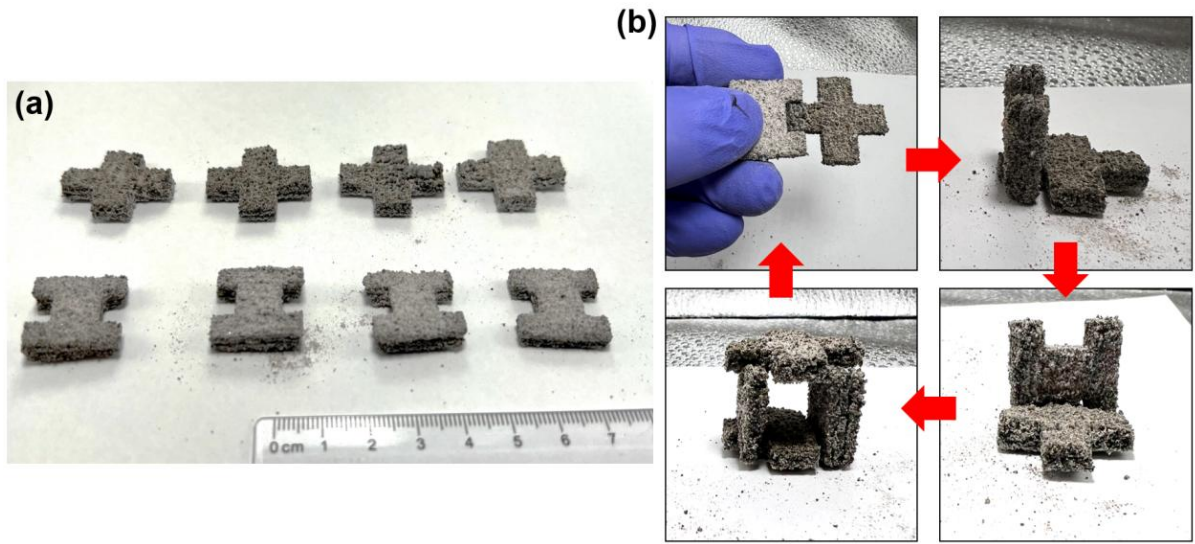

**Figure 4.** (a) Photographs of manufactured modular bricks; (b) Proof of concept demonstrations of reconfigurable assembly of bricks.

Two interconnectable structures at the centimeter scale are designed and manufactured, as shown in Figure 4 (a). One structure has an "H" shape, working as the female connection, and the other one has a "+" shape, working as the male connection unit. As a proof-of-concept, we show that these modular bricks can be connected to form different simple structures (Figure 3b), including a wall and a shielding unit. Different from previous selective laser sintering/melting work focused on small parts [9, 25], our reconfigurable brick designs can provide an alternative to using small parts to build large-scale constructions. Although these structures are at the centimeter scale, we envision that such modular bricks can be interconnected to produce larger structures and constructions. Due to the surface roughness and porosity, the connection realized by our current bricks is not as tight as Lego blocks. Yet, with further processing and structure optimization, the construction of complex and seamless habitation using modular bricks without additional materials is possible and promising.

Several challenges were identified during the reconfigurable assembly process. First, surface roughness impeded proper inter-brick contact, leading to insufficient mechanical engagement and tightness; this highlights the need for surface finish optimization to improve connection quality (especially tightness). Second, instances of delamination and structural failure were observed, primarily attributed to non-uniform sintering or partial melting; this underscores the need for a refined thermal processing approach to achieve fully densified and mechanically robust components. Generally, melting is preferred over sintering to obtain denser and smoother structures. Finally, the structural versatility of the current demonstration is limited by the geometry of the two available modular bricks, suggesting that additional brick designs are necessary to enable more complex and adaptable assemblies.

## 4. Conclusions

This research serves as proof of concept for future onsite manufacturing of small-scale reconfigurable bricks for large construction assembly, leveraging abundant lunar regolith and light energy. We demonstrated that with moderate laser processing energy density, different modular brick structures can be manufactured, and in the future assembled and reconfigured to different shapes. Although promising on the reconfigurable aspect, disadvantages were also observed, including delamination and anisotropic properties in the fabricated parts, and low mechanical strength, which requires further work to resolve for practical applications. Some key conclusions from this research are as follows.

(1) Under ambient conditions, a reasonable laser energy density (related to both power and scanning speed) is required to manufacture controllable structures.

(2) The mechanical properties of parts produced via this sintering method are highly anisotropic (the difference can be as high as 2 times). More research is needed to understand this anisotropicity and optimize the mechanical properties along different directions, especially to enhance the interlayer strength.

(3) Surface finish is critical for seamless connection between different modular bricks.

In spite of the limitations listed above, this research demonstrated the feasibility of reconfigurable modular bricks fabricated from lunar regolith using a light-based sintering approach. Future work should be conducted to understand the manufacturing process in a vacuum environment [14, 26], as well as the resulting materials in detail. It is envisioned that complex and large constructions can be developed using small, reconfigurable modular bricks.

**Acknowledgments**

Funding: This work was supported by the Honors College Undergraduate Research Grant and the Startup Package from the University of Arkansas.